\documentclass[submission,copyright,creativecommons]{eptcs}
\providecommand{\event}{WORDS 2011}
\usepackage{breakurl}             

\usepackage{amsfonts}
\usepackage{amsmath}
\usepackage{amssymb}
\usepackage{amsopn}
\usepackage{graphicx}
\usepackage{gastex}

\newtheorem{theorem}{Theorem}
\newtheorem{lemma}{Lemma}

\newtheorem{example}{Example}

\title{Infinite permutations vs. infinite words}
\author{Anna E. Frid
\institute{Sobolev Institute of Mathematics\\
4 Koptyug av., 630090, Novosibirsk, Russia}
\email{anna.e.frid@gmail.com}}

\def\titlerunning{Infinite permutations vs. infinite words}
\def\authorrunning{A. E. Frid}

\begin{document}
\maketitle

I am going to compare well-known properties of infinite words with those of infinite permutations, a new object studied since middle 2000s. Basically, it was Sergey Avgustinovich who invented this notion, although in an early study by Davis et al. \cite{degs} permutations appear in a very similar framework as early as in 1977. I am going to tell about periodicity of permutations, their complexity according to several definitions and their automatic properties, that is, about usual parameters of words, now extended to permutations and behaving sometimes similarly to those for words, sometimes not. Another series of results concerns permutations generated by infinite words and their properties. Although this direction of research is young, many people, including two other speakers of this meeting, have participated in it, and I believe that several more topics for further study are really promising.

\section{Definitions and examples}
Let ${\mathcal A}_{\mathbb N}$ be the set of all sequences of pairwise distinct reals defined
on $\mathbb N=\{0,1,2,\ldots\}$. Define an equivalence relation $\sim$ on ${\mathcal A}_S$
as follows: let $a,b$ be sequences from ${\mathcal A}_{\mathbb N}$, where $a=\{a_s\}_{s\in {\mathbb N}}$ and
$b=\{b_s\}_{s\in {\mathbb N}}$; then $a \sim b$ if and only if for all $s,r
\in {\mathbb N}$ the inequalities $a_s < a_r$ and $b_s<b_r$ hold or do not
hold simultaneously. An equivalence class from ${\mathcal A}_{\mathbb N} / \sim$ is
called an {\it \mbox{(${\mathbb N}$-)}permutation}, or a one-sided infinite permutation. If a permutation
$\alpha$ is realized by a sequence of reals $a$, that is, if the sequence $a$ belongs to the class $\alpha$, we
denote $\alpha=\overline{a}$. 

Similarly, we can consider a $\mathbb Z$-permutation, or an $S$-permutation, defined on an arbitrary subset $S$ of $\mathbb Z$.
In particular, a
$\{1,\ldots,n\}$-permutation always has a representative with all
values in $\{1,\ldots,n\}$, i.~e., can be identified with a usual
permutation from $S_n$.

In equivalent terms, an $S$-permutation can be considered as a linear ordering of the set $S$
which may differ from the ``natural'' one. It means that for $i,j \in S$, the natural
order between them corresponds to $i<j$ or $i>j$, while the ordering we intend to
define corresponds to $\alpha_i <\alpha_j$ or $\alpha_i >\alpha_j$. We shall also
use the symbols $\gamma_{ij}\in \{<,>\}$ meaning the relations between $\alpha_i$
and $\alpha_j$, so that by definition we have $\alpha_i \gamma_{ij} \alpha_j$ for
all $i \neq j$.

A {\em factor} of {\em length} $n$ of a finite or infinite permutation $\alpha_1\alpha_2 \cdots$ is any well-defined finite permutation $\alpha_{s+1}\alpha_{s+2}\cdots \alpha_{s+n}$ (considered as a $\{1,\ldots,n\}$-permutation).

\begin{example}{\rm
Let $\{a_i\}_{i=0}^{\infty}$ be the sequence defined by $a_n=(-1/2)^n$, and $\{b_i\}_{i=0}^{\infty}$ be the sequence defined by 
$b_i=1000+(-1)^n/n$. Then $\overline{a}=\overline{b}$; and we also can define the respective permutation $\alpha=\overline{a}=\overline{b}$ directly by the family of inequalities: for all $i,j\geq 0$ we have $\alpha_{2i}>\alpha_{2j+1}$, $\alpha_{2i}>\alpha_{2i+2}$, and $\alpha_{2j+1}<\alpha_{2j+3}$. Equivalently, the same family of inequalities can be written as $\gamma_{2i,2j+1}=>$, $\gamma_{2i,2i+2}=>$, and $\gamma_{2j+1,2j+3}=<$. It can be easily checked that these inequalities completely define the permutation, and that it is equal to $\overline{a}$ and to $\overline{b}$.

Note also that the permutation $\alpha$ cannot be represented by a sequence of integers since all its elements are sutuated between the first two of them.
}
\end{example}

\begin{example} \label{tm}
{\rm 
Let $w_{TM}$ be the Thue-Morse word, $w_{TM}=01101001\cdots$. 
Then we can associate with it the infinite permutation $\alpha_{TM}$ which is the order among the binary numbers $.01101001\cdots$, $.1101001\cdots$,
$.101001\cdots$, $.01001\cdots$, etc., equal to the shifts of the initial Thue-Morse binary number $.0110100110\cdots$. So, the first four values of
the permutation are ordered as $\alpha_3<\alpha_0<\alpha_2 < \alpha_1$; equivalently, the permutation $\alpha_0\alpha_1\alpha_2\alpha_3$ is equal to 2431.
In terms of the symbols $\gamma_{ij}\in \{<,>\}$ we have $\gamma_{01}=\gamma_{02}=<$ and $\gamma_{03}=\gamma_{12}=\gamma_{13}=\gamma_{23}=>$, etc.
}
\end{example}
\section{Periodicity}
A finite or infinite word $w=w_1w_2\cdots$ is called {\em $t$-periodic} if $w_{i}=w_{i+t}$ for all $i$ such that $w_{i+t}$ is well-defined. The number of $t$-periodic infinite words on a $q$-letter alphabet is equal to $q^t$.

Analogously, a finite or infinite permutation $\alpha=\alpha_1\alpha_2\cdots$ is called {\em $t$-periodic} if $\alpha_{i}<\alpha_{j}$ if and only if $\alpha_{i+t}<\alpha_{j+t}$, or, equivalently, if $\gamma_{ij}=\gamma_{i+t,j+t}$ for all $i$ and $j$. Surprisingly, for all $t>1$ the number of infinite $t$-periodic permutations is infinite \cite{ff}.

\begin{figure}
\centering \includegraphics[width=0.6\textwidth]{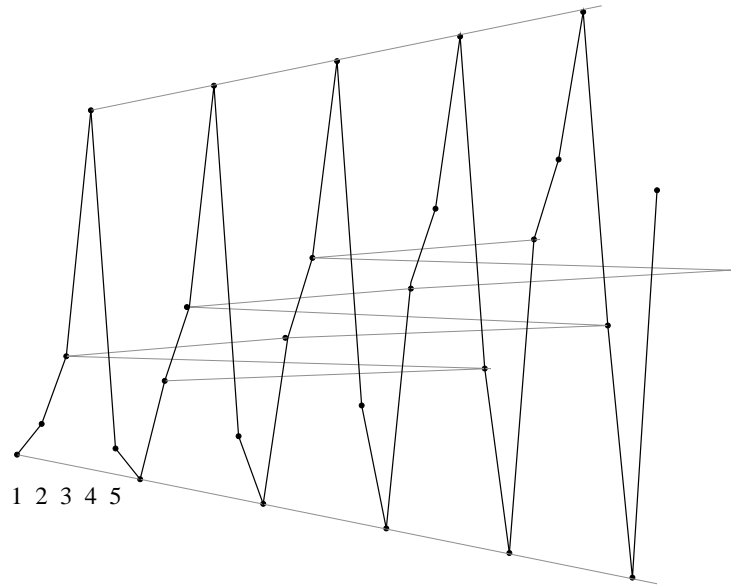}
\caption{A 5-periodic permutation.}
\label{5per}
\end{figure}

\begin{example}{\rm
Let us fix an $n$ and consider the the permutation $\alpha(n)$ with a representative sequence
\[a(n)=  1, \; 2n, \; 3, \; 2n+2, \ldots.\]
All permutations $\alpha(n)$ are 2-periodic and different.
}
\end{example}
In fact, for each $t>1$ the number of $t$-periodic permutations is countable. A typical 5-periodic permutation is depicted at Fig.~\ref{5per}. 

Now let us turn to finite permutations and investigate their Fine-Wilf properties. For words, the following famous theorem holds:
\begin{theorem}[Fine-Wilf]\label{t:fw_words}
If a word of length at least $p+q-(p,q)$ is $p$- and $q$-periodic, then it is $(p,q)$-periodic, and the length $p+q-(p,q)$ is the least possible. 
\end{theorem}

For permutations, the analogous result holds only partially.
\begin{theorem}
If a permutation $\alpha$ of length at least $p+q$ is $p$- and $q$-periodic with $(p,q)=1$, then $\alpha$ is 1-periodic, that is, monotonic. The length $p+q$ is the least possible. 
\end{theorem}

\begin{figure}\label{f:2}
\centering \includegraphics[width=0.35\textwidth]{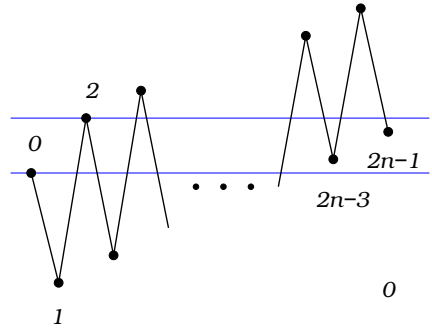}
\caption{An arbitrarily long 4- and 6-periodic but not 2-periodic permutation}
\end{figure}

The case when $p$ and $q$ are not coprime does not hold as it is shown at Fig.~\ref{f:2}.
However, the following fact can be considered as a version of Fine-Wilf theorem for general periods of permutations.

\begin{theorem}
Suppose that a permutation $\alpha$ of length $n$ is $p$- and $q$-periodic. Then each its factor of length at most   $n-p-q+2(p,q)+1$ is $(p,q)$-periodic.
\end{theorem}

It should be notices that with permutations, local periods have nothing common with the global period, as two consequtive permutations do not uniquely define their catenation. So, nothing similar to the critical factorization theorem for words can be stated for permutations.

\section{Low complexities}
Recall that the (subword) complexity of a word $w$ is the number $p_w(n)$ of its factors of length $n$. A word is ultimately periodic if and only if its complexity is ultimately constant; the lowest complexity of a non ultimately periodic word is $p_w(n)=n+1$, and the one-sided infinite words of this complexity are exactly Sturmian words.

For permutations, only the first statement holds. Define the {\em factor complexity} of a permutation $\alpha$ as the number $p_{\alpha}(n)$ of its factors of length $n$.

\begin{lemma}[\cite{ff}]
An $\mathbb N$-permutation is ultimately periodic if and only if its complexity is ultimately constant.
\end{lemma}

Clearly, the complexity of a permutation is bounded by $n!$, that is, it can grow faster than the complexity of a word on any finite alphabet. On the other hand, it can grow slower.

\begin{lemma}[\cite{ff}]
For each unbounded growing function $g(n)$ there exists a non-periodic $\mathbb N$-permutation $\alpha$ such that $p_{\alpha}(n)\leq g(n)$ for all sufficiently large $n$.
\end{lemma}

It is interesting that the situation with $\mathbb Z$-permutations is different.
\begin{lemma}[\cite{ff}]
For each non-periodic $\mathbb Z$-permutation $\alpha$ there exists a constant $C$, which can be arbitrarily large, such that $p_{\alpha}(n)\geq n-C$ for all $n$.
\end{lemma}

Now let us turn to maximal pattern complexity, introduced for words by Kamae and Zamboni \cite{kz}. Consider a window $S=\{0,t_1,\ldots,t_{n-1}\}$ of length $n$ and $S$-subwords $w_{m+S}=w_m w_{m+t_1}\cdots w_{m+t_n}$ of an infinite word $w$. The number of such words is called the $S$-complexity of $w$, and the maximum of $S$-complexities for windows $S$ of length $n$ is called the {\em maximal pattern complexity} $p^*_w(n)$ of $w$.

\begin{theorem}[\cite{kz}]
A word is ultimately periodic if and only if its maximal pattern complexity is ultimately constant. The least possible maximal pattern complexity of a non-periodic infinite word is $p^*_w(n)=2n$.
\end{theorem}
The words of complexity $p^*_w(n)=2n$ include Sturmian words and some Toeplitz words \cite{kz,kz2}; however, their characterization is not known. Fortunately, for permutations the situation is now clearer, as it was shown by Avgustinovich, Kamae, Salimov and the author in \cite{afks}. As above, we define the window $S$, the $S$-complexity of an infinite permutation $\alpha$ as the number of $S$-permutations $\alpha_{m+S}=\alpha_m \alpha_{m+t_1}\cdots \alpha_{m+t_n}$, and the maximal pattern complexity $p^*_{\alpha}(n)$ as the maximum of $S$-complexities for windows $S$ of length $n$.

\begin{theorem}[\cite{afks}]
An $\mathbb N$-permutation is ultimately periodic if and only if its maximal pattern complexity is ultimately constant. The least possible maximal pattern complexity of an $\mathbb N$-permutation is $p^*_w(n)=n$, and the permutations of complexity $n$ are exactly Sturmian permutations.
\end{theorem}
Here a Sturmian permutation $\alpha(w,x,y)$, where $w\in \{0,1\}^{\omega}$ is a Sturmian word, and $x$ and $y$ are rationally independent positive reals, is defined by a representative sequence $a$, where $a_0$ is arbitrary and 
\[a_{i+1}=\begin{cases} a_i+x, \mbox{~if~} w_i=0,\\
a_i-y, \mbox{~if~} w_i=1. \end{cases}\]

\section{Pattern avoidance}
On words, an occurrence of a {\em pattern}, that is, of a word on an alphabet of variables, is a finite word constructed from non-empty values of these variables in the given order. But all permutations of length one are equal, so that in some sense, no pattern is avoidable on permutations. However, we can restrict ourselves to values of variables of length at least two. The next question is how to define catenation: for example, which permutation is a square $XX$, 123 or better 1324? Avgustinovich, Kitaev, Pyatkin and Valyuzhenich \cite{akpv} chose the second version of the definition of a square and estimated the number of square-free permutations of length $n$ as $n^{n(1-\varepsilon_n)}$, where $\varepsilon_n \to 0$ with $n \to \infty$; they also investigated maximal permutations avoiding squares.

Other versions of the definition of patterns on permutations can be also interesting to consider.

\section{Permutations generated by words}
In the Example \ref{tm} above, we considered a way to define a permutation with the use of an infinite binary word. This way is general: given a word $w=w_0w_1\cdots$ on an alphabet $\{0,1,\ldots,q-1\}$, we can consider a permutation $\alpha$ generated by it as the permutation with a representative $a=\{a_i\}_{i=0}^\infty$, where $a_i=.w_iw_{i+1}w_{i+2}\cdots$ is a $q$-ary number. This definition is the most natural when the initial word is binary. A series of results on permutations generated by binary words has been obtained by M. A. Makarov \cite{mak1,mak2,mak3,mak4}. In particular, he found the maximal complexity of such a permutation \cite{mak1}, equal to 
\[p(n+1)=\sum_{t=1}^n \psi(t) 2^{n-t}=2^n(n-c+O(n 2^{-n/2})),\]
where 
\[\psi(t)=\sum_{d|t}\mu(t/d)2^d\]
is exactly the number of primitive binary words of length $t$. It is interesting that the complexity $p(n+1)$ is equal to the number of unary regular languages whose state complexity is at most $n$ \cite{dks}, and a bijection between these permutations and those languages is constructed in \cite{mak1}.

Makarov also investigated permutations generated by Sturmian words \cite{mak2}, the Thue-Morse word \cite{mak3} and the period doubling word \cite{mak4}. His research was continued by Widmer \cite{wid1,wid2} and  Valyuzhenich \cite{val1}. In particular, S. Widmer found the permutation complexity, that is, the complexity of the underlying permutation, of the Thue-Morse word \cite{wid1}; and A. Valyuzhenich extended the result to some its relatives \cite{val1}. All these results are rather technical and involve a study of several types of structures analogous to special subwords for words.

I would like to emphasize one of the results from \cite{mak3}, namely, the fact that the Thue-Morse permutation has another beautiful representative: the sequence 
\[0,1,\frac{1}{2},-\frac{1}{2},\frac{1}{4},-\frac{3}{4}, -\frac{1}{4},\frac{3}{4},\frac{1}{8},\ldots\]
of binary rationals generated by the morphism
\[\varphi: \begin{cases} x \to x/2, x/2-1 \mbox{~for~} x>0,\\
x \to x/2, x/2+1 \mbox{~for~} x\leq 1. \end{cases}\]
A sensible generalization of this ``morphic'' way of generating permutations would be interesting.

\section{Automatic properties}
\begin{figure}\label{f:tm}
\begin{center}
\unitlength 1mm
\begin{picture}(80,110)(-10,0)
\gasset{AHLength=4.0,AHlength=4,AHangle=9.09}
\gasset{Nw=10,Nh=10,Nmr=5}
\gasset{ELside=l}\gasset{loopCW=n}
\gasset{Nadjust=w,Nadjustdist=2,Nh=6,Nmr=1}
  \node(N1)(10,10){$0=0$}
  \node(N2)(50,10){$1=1$}
  \node(N3)(10,50){$0<1$}
  \node(N4)(50,50){$1>0$}
  \node(N5)(-30,90){$0>0$}
  \node(N6)(20,90){$1>1$}
  \node(N7)(40,90){$0<0$}
  \node(N8)(90,90){$1<1$}
  {\tiny
  \drawedge[curvedepth=4](N1,N2){$(1,1)$}
  \drawedge[curvedepth=4](N2,N1){$(1,1)$}
  \drawedge[curvedepth=4](N3,N4){$(1,1)$}
  \drawedge[curvedepth=4](N4,N3){$(1,1)$}
  \drawedge[curvedepth=4,ELpos=30](N1,N4){$(1,0)$}
  \drawedge[curvedepth=-4,ELside=r,ELpos=30](N2,N3){$(1,0)$}
  \drawloop[loopangle=180,ELside=r,ELpos=50](N1){$(0,0)$}
  \drawloop[loopangle=0,ELside=r,ELpos=50](N2){$(0,0)$}
  \drawloop[loopangle=180,ELside=r,ELpos=50](N3){$(0,0)$}
  \drawloop[loopangle=0,ELside=r,ELpos=50](N4){$(0,0)$}
  \drawedge(N1,N3){$(0,1)$}
  \drawedge[ELside=r](N2,N4){$(0,1)$}
  \imark[iangle=-90](N1)
  \rmark(N1)
\rmark(N2)
\rmark(N3)
\rmark(N4)

 \rmark(N5)
\rmark(N6)
\rmark(N7)
\rmark(N8)
 \drawloop[loopangle=90,ELside=r,ELpos=50](N5){$(0,0)$}
  \drawloop[loopangle=90,ELside=r,ELpos=50](N8){$(0,0)$}
  \drawloop[loopangle=90,ELside=r,ELpos=50](N6){$(0,0)$}
  \drawloop[loopangle=90,ELside=r,ELpos=50](N7){$(0,0)$}
 \drawedge[curvedepth=4](N5,N6){$(1,1)$}
  \drawedge[curvedepth=4](N6,N5){$(1,1)$}
   \drawedge[curvedepth=4](N7,N8){$(1,1)$}
  \drawedge[curvedepth=4](N8,N7){$(1,1)$}
  \drawedge[curvedepth=4](N3,N5){$(0,1)$}
  \drawedge[curvedepth=4](N5,N3){$(0,1)$}
   \drawedge[curvedepth=4](N3,N6){$(1,0)$}
  \drawedge[curvedepth=4](N6,N3){$(1,0)$}
    \drawedge[curvedepth=4](N4,N7){$(0,1)$}
  \drawedge[curvedepth=4](N7,N4){$(0,1)$}
   \drawedge[curvedepth=4](N4,N8){$(1,0)$}
  \drawedge[curvedepth=4](N8,N4){$(1,0)$}
   \drawedge[curvedepth=4](N6,N3){$(1,0)$}
    \drawedge[curvedepth=-40](N5,N4){$(1,0)$}
  \drawedge[curvedepth=-4](N6,N4){$(0,1)$}
   \drawedge[curvedepth=4](N7,N3){$(1,0)$}
  \drawedge[curvedepth=40](N8,N3){$(0,1)$}
}
\end{picture}
\end{center}
\caption{Automaton generating the Thue-Morse permutation}
\end{figure}
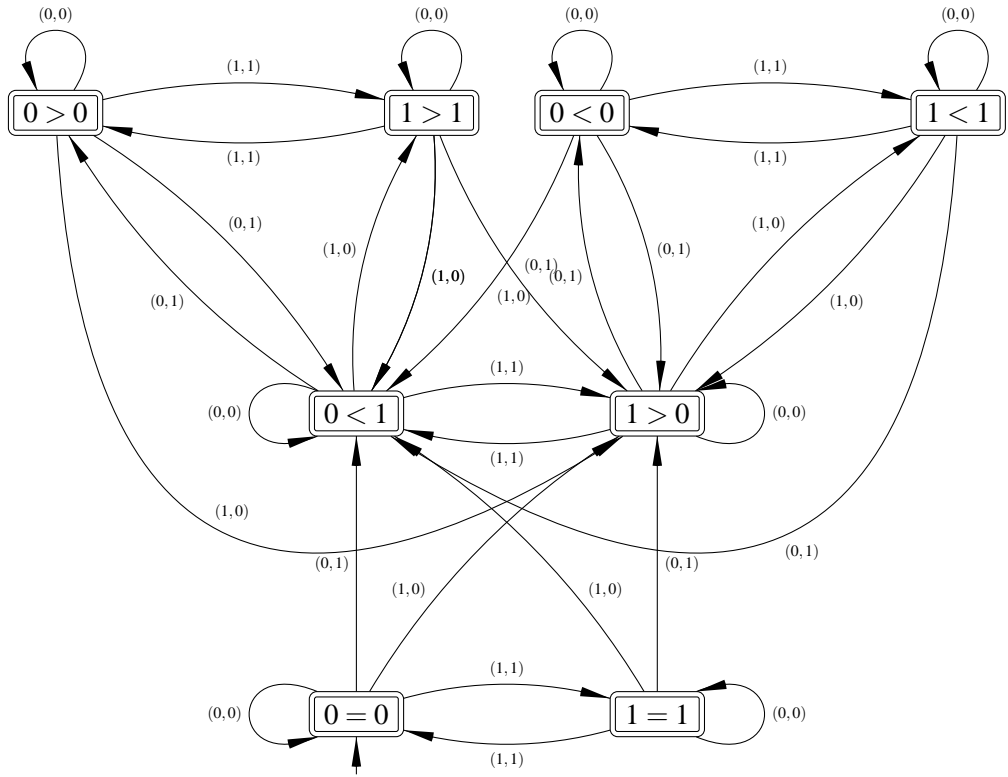

Several equivalent definitions of $k$-automatic words are well-known \cite{as}. For permutations, there is no such an equivalence; however, a permutation generated by a $k$-automatic word is $k$-automatic in the sense that the relation $\gamma_{ij}$ between any pair $\alpha_i$ and $\alpha_j$ of its elements can be found as the output of an automaton eating the pair of $k$-ary representations of $i$ and $j$ \cite{fz}. The number of states of the automaton given by our construction is huge; but Fig.~\ref{f:tm} shows a relatively small automaton generating the Thue-Morse permutation. The input of the automaton is pairs of digits of the binary representations of $i$ and $j$, starting from the most significant digit, with zeros in the beginning if necessary yo unify their lengths. The output is the relation $<$ or $>$ shown in the middle of each state; the lowest row of the automaton corresponds to the trivial case of $i=j$. 

\section{Conclusion}

Results concerning infinite permutations are in average more technical, less evident and sometimes more awkward than classical results for words. However, some of them look really beautiful. Several possible directions of further research are now outlined, and you are welcome to participate.

\bibliographystyle{eptcs}
\bibliography{anna}

\begin{thebibliography}{10}
\providecommand{\bibitemdeclare}[2]{}
\providecommand{\urlprefix}{Available at }
\providecommand{\url}[1]{\texttt{#1}}
\providecommand{\href}[2]{\texttt{#2}}
\providecommand{\urlalt}[2]{\href{#1}{#2}}
\providecommand{\doi}[1]{doi:\urlalt{http://dx.doi.org/#1}{#1}}
\providecommand{\bibinfo}[2]{#2}

\bibitemdeclare{book}{as}
\bibitem{as}
\bibinfo{author}{Jean-Paul Allouche} \& \bibinfo{author}{Jeffrey Shallit}
  (\bibinfo{year}{2003}): \emph{\bibinfo{title}{Automatic sequences --- Theory,
  applications, generalizations}}.
\newblock \bibinfo{publisher}{Cambridge University Press},
  \bibinfo{address}{Cambridge}, \doi{10.1017/CBO9780511546563}.

\bibitemdeclare{unpublished}{akpv}
\bibitem{akpv}
\bibinfo{author}{S.~V. Avgustinovich}, \bibinfo{author}{S.~Kitaev},
  \bibinfo{author}{A.~Pyatkin} \& \bibinfo{author}{A.~Valyuzhenich}:
  \emph{\bibinfo{title}{On square-free permutations}}.
\newblock \bibinfo{note}{Accepted to J. Autom. Lang. Comb.}

\bibitemdeclare{article}{afks}
\bibitem{afks}
\bibinfo{author}{S.V. Avgustinovich}, \bibinfo{author}{A.~Frid},
  \bibinfo{author}{T.~Kamae} \& \bibinfo{author}{P.~Salimov}
  (\bibinfo{year}{2011}): \emph{\bibinfo{title}{Infinite permutations of lowest
  maximal pattern complexity}}.
\newblock {\sl \bibinfo{journal}{Theoretical Computer Science}}
  \bibinfo{volume}{412}(\bibinfo{number}{27}), pp. \bibinfo{pages}{2911 --
  2921}, \doi{10.1016/j.tcs.2010.12.062}.

\bibitemdeclare{article}{degs}
\bibitem{degs}
\bibinfo{author}{J.~A. Davis}, \bibinfo{author}{R.~C. Entringer},
  \bibinfo{author}{R.~L. Graham} \& \bibinfo{author}{G.~J. Simmons}
  (\bibinfo{year}{1977/78}): \emph{\bibinfo{title}{On permutations containing
  no long arithmetic progressions}}.
\newblock {\sl \bibinfo{journal}{Acta Arith.}}
  \bibinfo{volume}{34}(\bibinfo{number}{1}), pp. \bibinfo{pages}{81--90}.

\bibitemdeclare{article}{dks}
\bibitem{dks}
\bibinfo{author}{Michael Domaratzki}, \bibinfo{author}{Derek Kisman} \&
  \bibinfo{author}{Jeffrey Shallit} (\bibinfo{year}{2002}):
  \emph{\bibinfo{title}{On the number of distinct languages accepted by finite
  automata with {$n$} states}}.
\newblock {\sl \bibinfo{journal}{J. Autom. Lang. Comb.}}
  \bibinfo{volume}{7}(\bibinfo{number}{4}), pp. \bibinfo{pages}{469--486}.

\bibitemdeclare{article}{ff}
\bibitem{ff}
\bibinfo{author}{D.~G. Fon-Der-Flaass} \& \bibinfo{author}{A.~E. Frid}
  (\bibinfo{year}{2007}): \emph{\bibinfo{title}{On periodicity and low
  complexity of infinite permutations}}.
\newblock {\sl \bibinfo{journal}{European J. Combin.}}
  \bibinfo{volume}{28}(\bibinfo{number}{8}), pp. \bibinfo{pages}{2106--2114},
  \doi{10.1016/j.ejc.2007.04.017}.

\bibitemdeclare{unpublished}{fz}
\bibitem{fz}
\bibinfo{author}{A.~Frid} \& \bibinfo{author}{L.~Zamboni}:
  \emph{\bibinfo{title}{On automatic infinite permutations}}.
\newblock \bibinfo{note}{Accepted to RAIRO -- Theoretical Informatics and
  Applications}.

\bibitemdeclare{article}{kz2}
\bibitem{kz2}
\bibinfo{author}{Teturo Kamae} \& \bibinfo{author}{Luca Zamboni}
  (\bibinfo{year}{2002}): \emph{\bibinfo{title}{Maximal pattern complexity for
  discrete systems}}.
\newblock {\sl \bibinfo{journal}{Ergodic Theory Dynam. Systems}}
  \bibinfo{volume}{22}(\bibinfo{number}{4}), pp. \bibinfo{pages}{1201--1214},
  \doi{10.1017/S0143385702000585}.

\bibitemdeclare{article}{kz}
\bibitem{kz}
\bibinfo{author}{Teturo Kamae} \& \bibinfo{author}{Luca Zamboni}
  (\bibinfo{year}{2002}): \emph{\bibinfo{title}{Sequence entropy and the
  maximal pattern complexity of infinite words}}.
\newblock {\sl \bibinfo{journal}{Ergodic Theory Dynam. Systems}}
  \bibinfo{volume}{22}(\bibinfo{number}{4}), pp. \bibinfo{pages}{1191--1199},
  \doi{10.1017/S0143385702000585}.

\bibitemdeclare{article}{mak4}
\bibitem{mak4}
\bibinfo{author}{M.~Makarov} (\bibinfo{year}{2010}): \emph{\bibinfo{title}{On
  the infinite permutation generated by the period doubling word}}.
\newblock {\sl \bibinfo{journal}{European J. Combin.}}
  \bibinfo{volume}{31}(\bibinfo{number}{1}), pp. \bibinfo{pages}{368--378},
  \doi{10.1016/j.ejc.2009.03.038}.

\bibitemdeclare{article}{mak1}
\bibitem{mak1}
\bibinfo{author}{M.~A. Makarov} (\bibinfo{year}{2006}):
  \emph{\bibinfo{title}{On permutations generated by infinite binary words}}.
\newblock {\sl \bibinfo{journal}{Sib. \`Elektron. Mat. Izv.}}
  \bibinfo{volume}{3}, pp. \bibinfo{pages}{304--311 (electronic)}.

\bibitemdeclare{article}{mak3}
\bibitem{mak3}
\bibinfo{author}{M.~A. Makarov} (\bibinfo{year}{2009}):
  \emph{\bibinfo{title}{On an infinite permutation similar to the
  {T}hue-{M}orse word}}.
\newblock {\sl \bibinfo{journal}{Discrete Math.}}
  \bibinfo{volume}{309}(\bibinfo{number}{23-24}), pp.
  \bibinfo{pages}{6641--6643}, \doi{10.1016/j.disc.2009.06.030}.

\bibitemdeclare{article}{mak2}
\bibitem{mak2}
\bibinfo{author}{M.~A. Makarov} (\bibinfo{year}{2009}):
  \emph{\bibinfo{title}{On permutations generated by {S}turmian words}}.
\newblock {\sl \bibinfo{journal}{Sibirsk. Mat. Zh.}}
  \bibinfo{volume}{50}(\bibinfo{number}{4}), pp. \bibinfo{pages}{850--857},
  \doi{10.1007/s11202-009-0076-6}.

\bibitemdeclare{unpublished}{val1}
\bibitem{val1}
\bibinfo{author}{A.~Valyuzhenich} (\bibinfo{year}{2011}):
  \emph{\bibinfo{title}{Permutation complexity of the fixed points of some
  uniform binary morphisms}}.
\newblock \bibinfo{note}{To appear in Proceedings of WORDS 2011}.

\bibitemdeclare{unpublished}{wid2}
\bibitem{wid2}
\bibinfo{author}{A.~Widmer} (\bibinfo{year}{2011}):
  \emph{\bibinfo{title}{Permutation complexity related to the letter doubling
  map}}.
\newblock \bibinfo{note}{To appear in Proceedings of WORDS 2011}.

\bibitemdeclare{article}{wid1}
\bibitem{wid1}
\bibinfo{author}{Steven Widmer} (\bibinfo{year}{2011}):
  \emph{\bibinfo{title}{Permutation complexity of the Thue-Morse word}}.
\newblock {\sl \bibinfo{journal}{Advances in Applied Mathematics}}
  \bibinfo{volume}{47}(\bibinfo{number}{2}), pp. \bibinfo{pages}{309 -- 329},
  \doi{10.1016/j.aam.2010.08.002}.

\end{thebibliography}

\end{document}